\documentclass[14p,prl,final,10pt,twocolumn]{revtex4}
\usepackage{amssymb}
\usepackage{epsfig,amsopn}

\input{tcilatex}
\begin{document}

\title{Correlated coherent oscillations in coupled semiconductor charge
qubits}
\author{Gou Shinkai$^{1,2}$}
\author{Toshiaki Hayashi$^{1}$}
\author{Takeshi Ota$^{1}$}
\author{Toshimasa Fujisawa$^{1,2}$}
\affiliation{$^{1}$NTT Basic Research Laboratories, NTT Corporation, 3-1
Morinosato-Wakamiya, Atsugi, 243-0198, Japan}
\affiliation{$^{2}$Department of Physics, Tokyo Institute of Technology, 2-12-1 Ookayama,
Meguro-ku, Tokyo, 152-8551, Japan}

\begin{abstract}
We study coherent dynamics of two spatially separated electrons in a coupled
semiconductor double quantum dot (DQD). Coherent oscillations in one DQD are
strongly influenced by electronic states of the other DQD, or the two
electrons simultaneously tunnel in a correlated manner. The observed
coherent oscillations are interpreted as various two-qubit operations. The
results encourage searching quantum entanglement in electronic devices.
\end{abstract}

\date{\today }
\maketitle

\FRAME{ftbhFU}{8.2cm}{7.3785cm}{0pt}{\Qcb{(Color) (a) Colored SEM image of
the control device (blue and black respectively for unetched and etched
surface, and gold for metal gates). The circles represent quantum dots in
DQD1 and DQD2. All measurements were performed at electron temperature of
100 mK at magnetic field of 0.6 T. The energy offsets, $\protect\varepsilon %
_{1}$ and $\protect\varepsilon _{2}$, and tunneling coupling, $\Delta _{1}$
and $\Delta _{2}$, were independently controlled by changing some gate
voltages simultaneously to compensate for electrostatic crosstalk. Tunneling
rates are $\Gamma _{L1}\sim \Gamma _{R1}\sim \Gamma _{L2}\sim \Gamma
_{R2}\sim $ 1 GHz for the left (L) and right (R) barrier of the first (1)
and second (2) DQD. (b) and (c) Charge diagram of the ground state at $%
\Delta _{1}=\Delta _{2}\sim 0$ in (b) and of the first excited state at $%
\Delta _{1}$ = 13 $\protect\mu $eV, $\Delta _{2}$ = 25 $\protect\mu $eV and $%
J$ = 25 $\protect\mu $eV in (c). Colors represent the charge state, LL
(cyan), LR (magenta), RL (yellow) and RR (white). The resonant conditions
are indicated by solid and dashed lines. Energy diagrams at some points ($%
\rhd \lhd \ominus \oplus $) for CROTs, SWAP, and FLIP operations are shown
in the insets. (d) Electrochemical potential of DQD1 and DQD2 in the steady
state for initialization at large bias (Ini1 and Ini2), in the coherent
evolution at zero bias (Evo1 and Evo2), and in the measurement at large bias
(Meas1 and Meas2). A schematic of the voltage pulse is shown at the top. }}{%
}{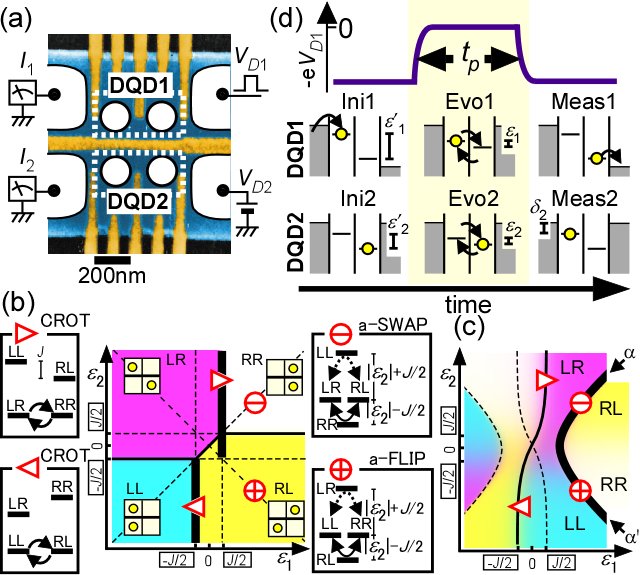}{\special{language "Scientific Word";type
"GRAPHIC";maintain-aspect-ratio TRUE;display "ICON";valid_file "F";width
8.2cm;height 7.3785cm;depth 0pt;original-width 4.4529in;original-height
3.6201in;cropleft "0";croptop "1";cropright "1";cropbottom "0";filename
'fig1.jpg';file-properties "XNPEU";}}

Quantum coherence of a single particle and a few particles (qubits) has been
investigated in various systems. Two-qubit unitary operations are key
ingredients for performing quantum algorithms and correlating multiple
qubits \cite{BookNielsenChuang}. Typical operations, such as
controlled-rotation (CROT), which rotates the target qubit state
conditionally on the control qubit state, and SWAP, which swaps quantum
states of the two qubits, have been demonstrated. However usually one type
of operations is realized depending on the type of coupling (Ising,
Heisenberg, etc.) \cite%
{ScienceGershenfeld,NatureYamamoto,PRLBrune,SciencePetta}. Although other
operations can in principle be designed in combination with some one-qubit
operations, it is not a simple task to generate different operations under
the same coupling. Simple sequences for shorter operation time or a smaller
number of steps have been desired to maintain the coherency of the system 
\cite{QICFowler,PRLBacon,PRACarliniTimeOptimal}. In this paper, we report
coherent dynamics for realizing multiple two-qubit operations. Artificial
and tunable charge qubits fabricated in a semiconductor nanostructure are
suitable for the purpose, since various parameters can be controlled by
external gate voltages. In the case of Ising (electrostatic dipole) coupling
between charge qubits fabricated in two double quantum dots (DQDs), coherent
oscillations of an electron in the first qubit can be controlled by the
second qubit's state (CROT operations) \cite{PRLBarenco,NanotechFedichkin}.
We find that, even under the same coupling, two spatially separated
electrons in the two DQDs change their locations coherently and collectively
(the correlated coherent oscillations). These coherent dynamics can be used
to design CROT, SWAP and other quantum operations, each in a single step.

We consider a system of two charge qubits, each of which possesses an excess
electron in an orbital of the left or the right dot, $|L\rangle _{i}$ or $%
|R\rangle _{i}$, of $i$-th semiconductor DQD ($i=1,2$) \cite%
{NatureOosterkamp,PRLHayashi,RMP03-Wilfred-DQD}. The electrostatic coupling
between the two qubits in the parallel geometry in Fig. 1(a) stabilizes
anti-parallel configurations, $|LR\rangle $ ($\equiv |L\rangle _{1}|R\rangle
_{2}$) and $|RL\rangle $, rather than parallel ones, $|LL\rangle $ and $%
|RR\rangle $. The coupling is expressed as an Ising-type Hamiltonian $\frac{J%
}{4}\sigma _{z}^{(1)}\otimes \sigma _{z}^{(2)}$, where $\sigma _{x,y,z}^{(i)}
$ are the Pauli matrices of the $i$-th qubit and $J$ ($>0$) represents the
strength of the coupling. The two-qubit system can be described by the
Hamiltonian%
\[
H_{2q}=\tfrac{1}{2}\sum\limits_{i}\left( \varepsilon _{i}\sigma
_{z}^{(i)}+\Delta _{i}\sigma _{x}^{(i)}\right) +\frac{J}{4}\sigma
_{z}^{(1)}\otimes \sigma _{z}^{(2)},
\]%
where the first term describes the energy offset $\varepsilon _{i}$ of the
single-qubit states and the tunneling coupling energy $\Delta _{i}$ of the $i
$-th qubit. A similar Hamiltonian can be found in various physical systems,
such as superconducting qubits \cite{NatureYamamoto} and ultra-cold atoms in
an optical lattice \cite%
{NatureAnderliniOpticalLattice,NatureFollingOpticalLattice}. Therefore, the
following arguments can also be applied to those systems.

Figure 1(b) illustrates the charge configuration of the ground state in the $%
\varepsilon _{1}-\varepsilon _{2}$ plane in the small but finite tunneling ($%
\Delta _{1}=\Delta _{2}\ll J$). The boundaries separating different charge
states represent the resonant conditions, where some important quantum
operations are expected. CROT operations are based on the coherent
transitions of a target qubit controlled by a control qubit \cite{PRLBarenco}%
. For example, a transition between $|LR\rangle $ and $|RR\rangle $ is
understood as rotating the state of the first qubit when the second qubit
state is $|R\rangle _{2}$. This transition is expected at $\varepsilon
_{1}=J/2$ [the right vertical line in Fig. 1(b)] and schematically shown in
the energy diagram of the upper-left inset. Here, $|LL\rangle $ and $%
|RL\rangle $ should be out of resonance (separated by $J$) for the CROT
operation. Another CROT operation for the transition between $|LL\rangle $
and $|RL\rangle $ appears at $\varepsilon _{1}=-J/2$ [the left vertical line
in Fig. 1(b) and corresponding energy diagram in the lower-left inset].
Similarly, CROT operations when the second qubit is the target are expected
at $\varepsilon _{2}=\pm J/2$ (the horizontal lines).

In contrast, the transition between $|LR\rangle $ and $|RL\rangle $ can not
be expected as a first-order process, and requires simultaneous tunneling of
two electrons via another state $|RR\rangle $ (or $|LL\rangle $) as
illustrated by the solid (dashed) arrow in the upper-right inset of Fig.
1(b). This second-order process takes place when $\Delta _{1}$ and $\Delta
_{2}$ are nonzero. When the first-order transitions are well suppressed ($%
||\varepsilon |\pm J/2|\gg \Delta _{1},\Delta _{2}$), the second-order
process dominates at $\varepsilon =\varepsilon _{1}=\varepsilon _{2}$ [on
the diagonal line $\ominus $ in Fig. 1(b)]. In this condition, eigenstates
of $H_{2q}$ are approximated to $\frac{1}{\sqrt{2}}\left( |LR\rangle \pm
|RL\rangle \right) $, $|RR\rangle $ and $|LL\rangle $, and the energy gap
between $\frac{1}{\sqrt{2}}\left( |LR\rangle \pm |RL\rangle \right) $ is\
given by $\Omega =\frac{1}{2}\Delta ^{2}J/|\varepsilon ^{2}-J^{2}/4|$ for
equal coupling $\Delta \equiv \Delta _{1}=\Delta _{2}$. This determines the
transition frequency between $|LR\rangle $ and $|RL\rangle $, and the
approximate SWAP operation is expected in a single step of the half period.
Approximate $\sqrt{\text{SWAP}}$ operation of the quarter period is useful
in preparation for correlated states $\frac{1}{\sqrt{2}}\left( |LR\rangle
\pm i|RL\rangle \right) $.

Similarly, another second-order transition between $|LL\rangle $ and $%
|RR\rangle $ [See energy diagram in the lower-right inset of Fig. 1(b)] is
allowed at $\varepsilon =\varepsilon _{1}=-\varepsilon _{2}$ [on the
diagonal line $\oplus $ in Fig. 1(b)], which can be used to flip the total
charge polarization of the two-qubit system in a single step (We call this
process FLIP). The approximate $\sqrt{\text{FLIP}}$ operation is useful in
preparation of correlated states $\frac{1}{\sqrt{2}}\left( |LL\rangle \pm
i|RR\rangle \right) $. In this way, multiple two-qubit operations (CROTs,
SWAP and FLIP) can be performed each in a single step.

We implemented a two-qubit system by integrating two sets of DQDs in a
GaAs/AlGaAs heterostructure, as shown in the scanning electron micrograph
(SEM) in Fig. 1(a) \cite{APLShinkai}. The two DQDs with individual source
and drain electrodes are electrically isolated, and thus independent
currents, $I_{1}$ and $I_{2}$, can be measured simultaneously. All qubit
parameters can be controlled by 11 gate voltages. We have already confirmed
the electrostatic coupling between the DQDs from the resonant tunneling
characteristics, in which the resonant tunneling of the first DQD is
switched by the charge state of the second DQD \cite{APLShinkai}.

The pulse sequence consists of the following three steps, similar to our
previous one-qubit experiment \cite{PRLHayashi}. First, the system is
initialized by setting DQD1 in the dissipative single-electron tunneling
regime at $V_{D1}$ = 700 $\mu $V, as illustrated in Ini1 of Fig. 1(d). An
electron is prepared in the left dot with a high probability (long dwell
time in the left dot) \cite{JVSTFujisawa}. This initialization works at any
energy offset $\varepsilon _{1}^{\prime }$ (Hereafter, the prime is used for
the biased situation). On the other hand, DQD2 is kept in the Coulomb
blockade region with a small bias $V_{D2}\sim $ 10 $\mu $V (See Ini2), and
thus the second qubit is relaxed in the steady state. Therefore, the system
is initialized in $|LL\rangle $ at $\varepsilon _{2}^{\prime }<-J/2$ or $%
|LR\rangle $ at $\varepsilon _{2}^{\prime }>-J/2$. Then, the system is
suddenly brought into the Coulomb blockade condition by applying a square
(negative) voltage pulse ($V_{D1}$ = 0) with a rise time of about 0.1 ns for
a period of $t_{p}$ = 0.08 -- 2 ns, where coherent time evolution is
expected (See Evo1 and Evo2 illustrated for the SWAP action). The readout of
the final state is performed by restoring the large bias ($V_{D1}$ = 700 $%
\mu $V), where the electron in the right dot escapes to the drain and
contributes to the current (Meas1). The above sequence is repeated at 100
MHz to obtain measurable current. We obtained the net electron numbers
flowing per pulse, $N_{P1}$ and $N_{P2}$, for each DQD by using lock-in
amplifiers \cite{PRLHayashi}. Inelastic tunneling during the initialization
period (10 ns $-t_{p}$) gives a background artifact proportional to $t_{p}$
[-0.12 $\sim $ -0.16$\times t_{p}$(ns)], which was partially subtracted to
highlight the coherent oscillations.

\FRAME{fbFU}{7.8002cm}{9.7925cm}{0pt}{\Qcb{(Color) (a) and (b) First-order
coherent oscillations (CROTs) for the initial state $|LR\rangle $ in (a) and
for $|LL\rangle $ in (b). (c) to (e) $N_{P1}$ in the $\protect\varepsilon %
_{1}-\protect\varepsilon _{2}$ plane obtained at a fixed $t_{p}$ = 0.25 ns ($%
\protect\pi $-pulse for the CROTs). The energy offsets in the initialization
period ($\protect\varepsilon _{1}^{\prime }$ and $\protect\varepsilon %
_{2}^{\prime }$) are shown in the right and top scales [$\protect\varepsilon %
_{1}^{\prime }\simeq \protect\varepsilon _{1}-$ 50 $\protect\mu $eV and $%
\protect\varepsilon _{2}^{\prime }\simeq \protect\varepsilon _{2}-$ 20 $%
\protect\mu $eV]. (f) Correlated coherent oscillations at $\protect%
\varepsilon _{1}\sim $ 25 $\protect\mu $eV along the solid line in (e). (g)
Density matrix simulation of (f). (h) Typical $N_{p1}\left( t_{p}\right) $
traces; the first-order [ $\protect\varepsilon _{1}=-J/2$ of (b)] and
correlated coherent oscillations ($\protect\varepsilon _{2}\sim $ 10 $%
\protect\mu $eV).}}{}{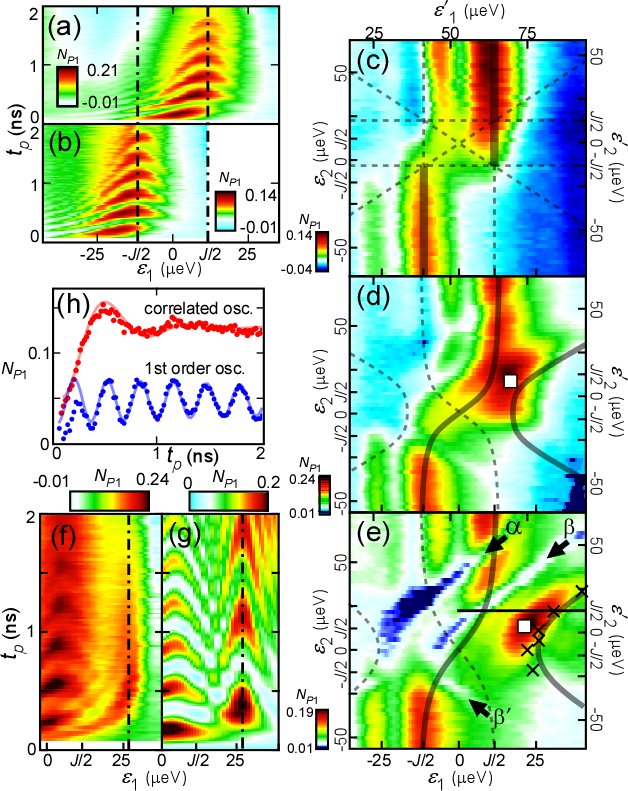}{\special{language "Scientific Word";type
"GRAPHIC";maintain-aspect-ratio TRUE;display "ICON";valid_file "F";width
7.8002cm;height 9.7925cm;depth 0pt;original-width 3in;original-height
5.8202in;cropleft "0";croptop "1";cropright "1";cropbottom "0";filename
'fig2.jpg';file-properties "XNPEU";}}

We first demonstrate the CROT operation of DQD1 using DQD2 as a control
qubit. Figure 2(a) shows the coherent oscillations of DQD1 starting from the
initial state $|LR\rangle $ prepared at $\varepsilon _{2}^{\prime }$ = +35 $%
\mu $eV. The overall behavior of the oscillations, including the dependence
of the period, amplitude, and decoherence time on the detuning from the
resonance ($\varepsilon _{1}=J/2$), is consistent with our previous study on
a single qubit \cite{PRLHayashi} and with the two-qubit simulation, from
which we obtained $\Delta _{1}$ = 13 $\mu $eV. When the initial state is
prepared to be $|LL\rangle $ at $\varepsilon _{2}^{\prime }$ = -60 $\mu $eV,
similar oscillations are observed as shown in Fig. 2(b) but with a resonance
appearing at a significantly different value of $\varepsilon _{1}$.
Crossover between the two oscillations is depicted in Fig. 2(c), where $%
N_{p1}$ measured at $t_{p}$ = 0.25 ns (corresponding to $\pi $-pulse on
resonance) is plotted in the $\varepsilon _{1}-\varepsilon _{2}$ plane. The
resonant conditions discussed in Fig. 1(b)\ are superimposed in Fig. 2(c)
for clarity. The vertical patterns of the interference fringes are
horizontally shifted discontinuously when the initial state is altered from $%
|LR\rangle $ in the upper region to $|LL\rangle $ in the lower region. The
horizontal shift is a measure of the electrostatic coupling $J$ = 25 $\mu $%
eV. The large $J$ as compared to $\Delta _{1}$ (13 $\mu $eV) ensures
reasonable controllability of CROT.

The above CROT experiments were performed at small tunneling coupling of
DQD2 ($\Delta _{2}\sim $ 3 $\mu $eV estimated from an independent
measurement). As $\Delta _{2}$ is increased while keeping $\Delta _{1}$
constant, the second-order transitions appear as additional features
(labeled by $\square $) elongated in the upper-right direction in Fig. 2(d)
for $\Delta _{2}\sim $ 15 $\mu $eV and in Fig. 2(e) for $\Delta _{2}\sim $
25 $\mu $eV (determined from the following analysis). The time evolution of $%
N_{P1}\left( t_{p}\right) $ is also investigated at various $\varepsilon
_{1} $ and $\varepsilon _{2}$. Figure 2(f) shows the $\varepsilon _{1}$
dependence of $N_{P1}\left( t_{p}\right) $ at $\varepsilon _{2}=$ 10 $\mu $%
eV [the dot-dashed line in Fig. 2(e)]. Two types of oscillations are
resolved: one at $\varepsilon _{1}\sim 0$ with the frequency comparable to
that of the first-order process and the other at $\varepsilon _{1}\sim 25$ $%
\mu $eV with a lower frequency of 1.3 GHz. Figure 2(h) shows typical $%
N_{P1}\left( t_{p}\right) $ plots of the two cases. We identify the slower
oscillations to be correlated dynamics of the two qubits from the following
analysis.

The resonant conditions $\left( \varepsilon _{1},\varepsilon _{2}\right) $
were experimentally determined at the maximum $N_{P1}$ for sufficiently long
pulse ($t_{p}\sim $ 1 ns) [for example, the dot-dashed line in Fig. 2(f)],
and are plotted by crosses ($\times $) in Fig 2(e). These conditions can be
reproduced just by considering the eigenstates of $H_{2q}$. The resonant
conditions were numerically derived for the minimum energy gap between the
eigenstates, and are shown by solid and dashed gray lines in Fig 2(e) with a
fitted parameter $\Delta _{2}=$ 25 $\mu $eV. One of these line fits well
with the experimental resonant conditions. The same resonance lines together
with the charge state (shown by colors) of the first excited state are shown
in Fig. 1(c) with the same parameters. As recognized by the neighboring
charge states, the condition of interest (the thick line $\alpha -\alpha
\prime $) represents the resonance of $|LR\rangle $ and $|RL\rangle $ (SWAP)
for $\varepsilon _{2}\gg J/2$, that of $|LL\rangle $ and $|RR\rangle $
(FLIP) for $\varepsilon _{2}\ll -J/2$, and that of superpositions of four
bases for $\varepsilon _{2}\sim 0$. Namely, the straight resonant conditions
in Fig. 1(b) for $\Delta _{1}=\Delta _{2}$ are curved as in Fig. 1(c) for
unequal coupling ($\Delta _{1}<\Delta _{2}$). The observed coherent dynamics
at $\varepsilon _{2}\sim 0$ involves complicated superposition of four
two-qubit bases, and thus may be called correlated coherent oscillations
rather than simpler SWAP or FLIP operations.

We also performed density-matrix simulations using the standard Lindblad
master equation \cite{JVSTFujisawa,PhysRepBrandes}. In addition to the
coherent processes described by $H_{2q}$, incoherent tunneling transitions
to/from the source and drain electrodes and spontaneous phonon emission in
each DQD were included with realistic parameters. To reproduce the
experimental pulse sequence, the time-dependent reduced density matrix $\rho
\left( t\right) $ in the Coulomb blockade region (at $V_{D1}=$ 0) was
numerically calculated from the initial state $\rho _{0}$ prepared in the
transport region (at finite $V_{D1}$). Expected tunneling electrons
(measurement outcome) in the subsequent transport region after the pulse
length $t_{p}$ were evaluated. The simulated $N_{p1}$ is plotted in Fig.
2(g) by using the same parameter in Fig. 2(f). Although the correlated
oscillations are significantly degraded by other decoherence mechanisms
(charge noise, finite rise-time of the pulse, etc.), the overall oscillation
characteristics (fast and slow oscillations, $\varepsilon _{1}$-dependent
period around the resonance) are well reproduced in the simulations.

The correlated tunneling can be evidenced by measuring the signal in the
DQD2. However, no readout signal in $N_{P2}$ for DQD2 was obtained in the
above measurements, since they were performed in the Coulomb blockade
condition of DQD2 [See diagrams in Fig. 1(d)]. We confirmed a small but
finite \textit{negative} signal in $N_{P2}$ (but \textit{positive} in $%
N_{P1} $) when DQD2 was made closer to the conductive region by adjusting
the electrochemical potential $\delta _{2}$ close to zero (data not shown).
This indicates a fraction of SWAP action (transition from $|LR\rangle $ to $%
|RL\rangle $ produces negative current in DQD2). Since the second-order
coupling decreases with increasing $|\varepsilon _{1}|$, we could confirm
coherent oscillations in the limited range of $|\varepsilon _{1}|\lesssim $
25 $\mu $eV. Simpler SWAP and FLIP operations are expected at larger $%
|\varepsilon _{1}|$.

We examined such conditions from dc measurement. Figure 3 shows (a) $I_{1}$
at $V_{D1}$ = 700 $\mu $V and (b) $I_{2}$ at $V_{D2}$ = 10 $\mu $V.
First-order tunneling current in DQD1 and DQD2 is seen as broad vertical
lines [outside the plot range of (a)] and a horizontal line at around $%
\varepsilon _{2}^{\prime }=\pm J/2$ in (b). The second-order cotunneling is
simultaneously recorded in both currents and appears as very sharp current
peaks $\alpha $, $\beta $, and $\beta ^{\prime }$ running in the diagonal
directions. Peak $\alpha $ is associated with the two-qubit system of
interest, and $\beta -\beta ^{\prime }$ involves another orbital state (an
excited state) neglected in the model. The negative current correlation
(positive $I_{1}$ but negative $I_{2}$)\ appearing in the upper region ($%
\varepsilon _{2}^{\prime }>0$) is understood as the cotunneling transition
from the initial state $|LR\rangle $ to the final state $|RL\rangle $ at $%
\varepsilon _{1}^{\prime }\sim \varepsilon _{2}^{\prime }>J/2$ [See arrows
in the upper-right inset of Fig. 1(b)]. On the other hand, the positive
correlation (positive $I_{1}$ and $I_{2}$) appearing in the lower region ($%
\varepsilon _{2}^{\prime }<0$) of Fig. 3 arises from the transition from $%
|LL\rangle $ to $|RR\rangle $ at $\varepsilon _{1}^{\prime }\sim
-\varepsilon _{2}^{\prime }>J/2$ [See the lower-right inset of Fig. 1(b)].
The overall second-order tunneling peak constitutes the resonant condition $%
\alpha -\alpha \prime $ in Fig. 1(c). Although coherent oscillations were
confirmed only at $\varepsilon _{2}\sim 0$, the appearance of cotunneling
peak in the wide region in Fig. 3 suggests that coherent SWAP and FLIP
operations are feasible. \FRAME{fbFU}{6.9369cm}{4.2043cm}{0pt}{\Qcb{%
Correlated cotunneling current in simultaneous measurement of $I_{1}$ in (a)
and $I_{2}$ in (b) in the $\protect\varepsilon _{1}^{\prime }-\protect%
\varepsilon _{2}^{\prime }$ plane measured at $V_{D1}$ = 700 $\protect\mu $%
eV and $V_{D2}$ = 10 $\protect\mu $eV. Solid and dotted lines are current
traces at $\protect\varepsilon _{2}^{\prime }\sim $ 50 $\protect\mu $eV and
-10 $\protect\mu $eV, respectively. Second-order tunneling peaks $\protect%
\alpha $, $\protect\beta $, and $\protect\beta ^{\prime }$ apear at the same
marks in Fig. 2(e). }}{}{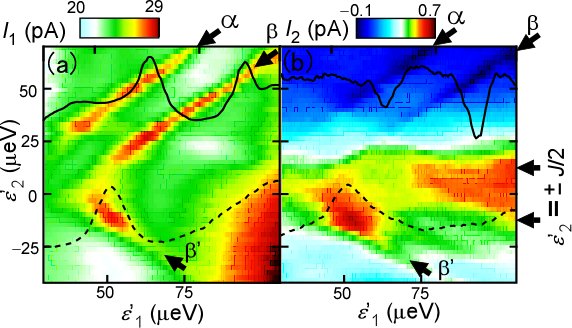}{\special{language "Scientific Word";type
"GRAPHIC";maintain-aspect-ratio TRUE;display "ICON";valid_file "F";width
6.9369cm;height 4.2043cm;depth 0pt;original-width 4.1139in;original-height
2.2338in;cropleft "0";croptop "1";cropright "1";cropbottom "0";filename
'fig3.jpg';file-properties "XNPEU";}}

Finally we'd like to make all data consistent. Data in Fig. 3 were taken at
the same condition as in Fig. 2(e) but in the absence of a pulse ($t_{p}$ =
0). Negative signals, where coherent dynamics vanishes, along the diagonal
lines marked by $\alpha $, $\beta $, and $\beta ^{\prime }$ in Fig. 2(e)
coincide with the cotunneling currents in Fig. 3, and are attributed to the
second-order cotunneling occurring in the initialization period. This spoils
the initialization and measurement scheme and thus the coherent dynamics
disappears. Such experimental details were all reproduced in the
density-matrix simulation (not shown).

In summary, we have investigated first- and second-order tunneling processes
by measuring pulse-induced current or current correlations. We expect that
multiple two-qubit operations (CROT, SWAP and FLIP) can be induced just by
changing parameters $\varepsilon _{1}$ and $\varepsilon _{2}$.
Interestingly, the two electrons in our device have no spatial overlap of
wavefunctions, and the expected quantum processes therefore allow us to
correlate two electrons (a few hundred nanometers apart) without their
touching each other. This will be an important test of quantum non-locality
in mesoscopic electron devices and encourages the study of charge-based
quantum information even under a constant coupling \cite%
{PhysRepBrandes,NatureNederEntangle,RMPAmico}.

This work was partially supported by SCOPE from the Ministry of Internal
Affairs and Communications of Japan, KAKENHI (19204033) from the JSPS, and
the GCOE program 'Nanoscience and Quantum Physics' at TokyoTech. We thank M.
Eto, Y. Kawano, K. Muraki, Y. Tokura, H. Yamaguchi for valuable comments.

\end{document}